\documentclass[]{spie}

\usepackage{amsmath,amsfonts,amssymb}
\usepackage{hyperref}
\usepackage{graphicx}
\usepackage{setspace}
\usepackage{tocloft}
\usepackage{easy-todo}
\usepackage{physics}

\renewcommand{\vec}{\mathbf}
\newcommand{\E}{\mathbb{E}}
\newcommand{\dare}{\text{dare}}
\newcommand{\diag}{\text{diag}}
\newcommand{\mat}[1]{\begin{bmatrix}#1\end{bmatrix}}

\title{Laboratory Demonstration of Optimal Identification and Control of Tip-Tilt Systems}

\author[a,b]{Aditya R. Sengupta}
\author[a]{Benjamin L. Gerard}
\author[a]{Daren Dillon}
\author[a]{Maaike van Kooten}
\author[a]{Donald Gavel}
\author[a]{Rebecca Jensen-Clem}

\affil[a]{Department of Astronomy and Astrophysics, University of California, Santa Cruz, United States}
\affil[b]{Department of Applied Mathematics and Theoretical Physics, University of Cambridge, United Kingdom}

\authorinfo{Further author information: (Send correspondence to A.S.)\\A.S.: E-mail: adityars@ucsc.edu, Telephone: 1 510 697 6549}

\cftpagenumbersoff{figure}
\cftpagenumbersoff{table} 

\begin{document}

\maketitle
\begin{abstract}
    We present the results of testing optimal linear-quadratic-Gaussian (LQG) control for tip and tilt Zernike wavefront modes on the SEAL (Santa cruz Extreme AO Lab) testbed. The controller employs a physics model conditioned by the expected tip/tilt power spectrum and vibration peaks. The model builds on similar implementations, such as that of the Gemini Planet Imager, by considering the effects of loop delays and the response of the control hardware.  Tests are being performed on SEAL using the Fast Atmospheric Self-coherent camera Technique (FAST), and being executed using a custom Python library to align optics, generate interaction matrices, and perform real-time control by combining controllers with simulated disturbance signals to be corrected. We have carried out open-loop data collection, characterizing the natural bench dynamics, and have shown a reduction in RMS wavefront error due to integrator control and LQG control.


\end{abstract}


\section{Introduction}


Adaptive optics (AO) systems work to correct wavefront errors in images captured by optical and near infrared telescopes, via a control loop that measures deviations with a wavefront sensor (WFS) and corrects them by applying commands to a deformable mirror (DM) in the loop. AO has been essential to recent developments in astronomy, from ground-based direct imaging and spectroscopy of exoplanets, to precision astrometry\cite{SgrAStar}. Within AO systems, correcting the wavefront tip and tilt modes at the plane of the DM, which corresponds to $x-y$ position deviations for the downstream focal plane point spread function, is critical in enabling high-quality AO performance, and can provide substantial benefits if optimized. 

A major example of the importance of tip-tilt correction is the Keck Planet Imager and Characterizer (KPIC)\cite{KPIC}, which aims to carry out direct imaging and spectroscopy of young exoplanets. It aims to couple light from the planet into an optical fiber so that the light can be transported to an extremely stable spectrograph, producing a high-resolution spectrum. This requires that the point-spread-function (PSF) of the planet be centered on the fiber tip. Any movement in tip-tilt will move the PSF off the fiber, resulting in a reduced amount of planet light being transmitted through the fiber and reducing the resolution of spectra. This process is especially sensitive to high-frequency motion in tip-tilt\cite{hottinger2019focal}. A control algorithm for tip-tilt that minimizes motion while damping high-frequency disturbance components is therefore critical for these scientific applications.

Optimal control solutions for tip-tilt generally involve using a recent history of noisy WFS measurements to predict and control for future states. In particular, the case of linear-quadratic-Gaussian (LQG) control\cite{LeRoux,Meimon,Sivo,Petit} has been extensively studied. Assuming stationarity of turbulence statistics, linear dynamics perturbed by independent and identically distributed (i.i.d.) Gaussian process noise, and measurements perturbed by i.i.d. Gaussian measurement noise, the LQG controller provides the optimal (minimum-variance) sequence of control inputs. Another solution is predictive wavefront control\cite{Guyon}, which involves using a recent history of noisy WFS measurements to predict and control for future states without a particular model for the driving process. LQG has the advantage of being computationally lightweight, involving only two relatively small matrix multiplications per time step.

In this paper, we present the results of experiments run on the UCSC SEAL (Santa Cruz Extreme AO Laboratory) testbed implementing LQG control for tip-tilt. We make use of the FAST (Fast Atmospheric Self-coherent camera Technique) method\cite{FAST18}, which is a focal plane WFS designed for exoplanet imaging science that enables tip/tilt measurement and correction with the coronagraphic science image (see appendix \ref{sec: fast_rev} for a review of FAST), and demonstrate via timing experiments that the software and hardware setup used is robust with arbitrarily computationally intensive control laws, and with a hardware-limited loop rate of 100 Hz. We discuss the mathematical foundation of LQG control adapted to this case, including system identification of disturbances and controller dynamics, and demonstrate improved control residuals as a result of this identification.


In section 2, we describe the mathematics behind LQG control for tip-tilt and present simulation results. In section 3, we describe the software and hardware setup, calibration steps, and characterization of the control loop timings for an arbitrary control law. In section 4, we present and discuss the results of bench testing, and discuss future work.

\section{Methods}

\newcommand{\obs}{\text{obs}}
\newcommand{\con}{\text{con}}

\subsection{Linear-quadratic-Gaussian control}

Optimal control refers to a strategy for applying control inputs to a dynamic system such that its state minimizes a cost function. In the case of adaptive optics, we consider linear discrete-time dynamic systems with states $\vec{x}_k$, measurements $\vec{y}_k$, and with the process and measurement perturbed by i.i.d. Gaussian noise $\vec{w}_k \sim \mathcal{N}(0, W), \vec{v}_k \sim \mathcal{N}(0, V)$ respectively, where $W, V$ are known covariance matrices.

\begin{align}
    \begin{split}
        \vec{x}_{k+1} &= A\vec{x}_k + B\vec{u}_k + \vec{w}_k\\
        \vec{y}_k &= C\vec{x}_k + D\vec{u}_k + \vec{v}_k
    \end{split}
\end{align}

LQG control depends on solving for the optimal observer (estimating the state given the noisy measurements) and controller (computing inputs based on the estimated state). These optimal expressions are derived in Appendix \ref{sec:lqg_rev}. Both the observer and controller problems come down to solving a certain quadratic matrix equation, known as the discrete algebraic Riccati equation (DARE). We numerically solve the DARE with a series of methods, since it can be ill-conditioned with certain models and having multiple methods makes the implementation more reliable. We first use the solver in the Slycot\cite{slycot} Python wrapper to the SLICOT numerical control library\cite{SLICOT}, then the \textit{scipy.linalg.solve\_discrete\_are} method\cite{scipy}, and finally numerical iteration, in which we repeatedly set $P$ equal to the right-hand side of~\ref{eq:dare}, and iterate until we reach a fixed point. We denote the solution to the DARE by $P = \dare(\dots)$, where the arguments are functions of the model matrices shown in Table \ref{tab:matrices}.

The solvability of the DARE results in the \textit{Kalman filter} equations for the observer,

\begin{subequations}
\begin{align}
        P_\obs &= \dare(A^\intercal, C^\intercal, W, V, \mathbf{0})\\
        K &= P_\obs C^\intercal (CP_\obs C^\intercal + V)^{-1}\\
        \label{eq:kalman_predict} \tilde{\vec{x}}_{k+1} &= A\hat{\vec{x}}_k + B\vec{u}_k\\
        \label{eq:kalman_update} \hat{\vec{x}}_{k+1} &= \tilde{\vec{x}}_{k+1} + K(\vec{y}_k - C\tilde{\vec{x}}_{k+1}),
\end{align}
\end{subequations}

and the \textit{linear-quadratic regulator} (LQR) equations for the controller:

\begin{subequations}
\begin{align}
    P_\con &= \dare(A, B, C^\intercal C, D^\intercal D, D^\intercal C)\\
    L &= -(D^\intercal D + B^\intercal P_\con B)^{-1}(D^\intercal C + B^\intercal P_\con A)\\
    \label{eq:lqr_control} \vec{u}_k &= L\vec{x}_k.
\end{align}
\end{subequations}

A key benefit of LQG control is that the gains $K$ and $L$ can be precomputed, meaning that after the model identification step, the only steps that need to be computed per timestep are the Kalman prediction (\ref{eq:kalman_predict}), the Kalman update step (\ref{eq:kalman_update}), and the LQR input computation (\ref{eq:lqr_control}).

\begin{figure}
    \centering
    \includegraphics[width=\textwidth]{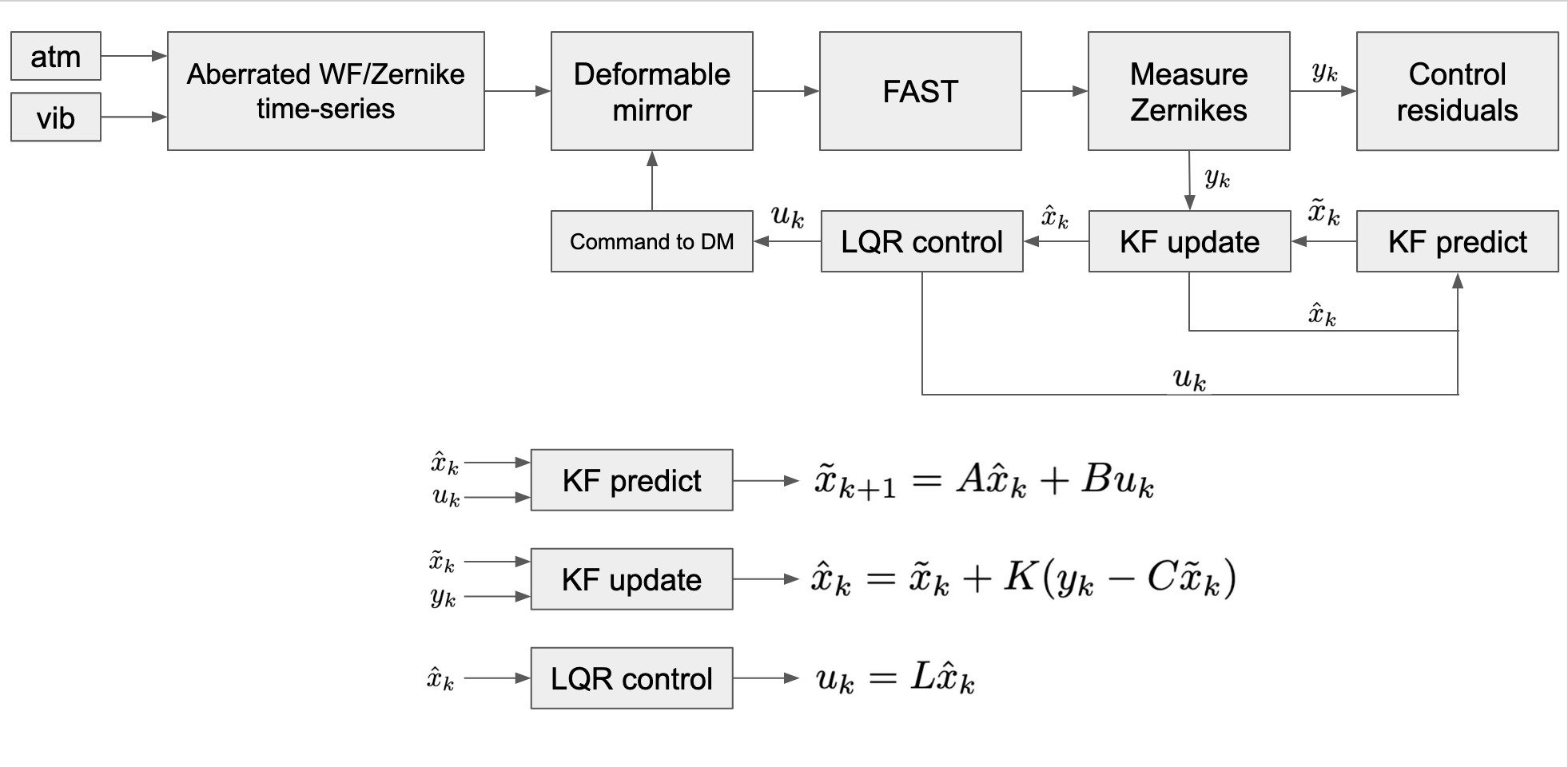}
    \caption{LQG control in an adaptive optics loop. The forward arm consists of the plant (the DM and FAST), and the backward arm contains the linear algebra steps required to implement LQG control in real time, given known values of $A, B, C, K, L$.}
    \label{fig:lqg_block}
\end{figure}

Figure~\ref{fig:lqg_block} depicts an overview of LQG control in an adaptive optics control loop, which shows that control requires only five matrix multiplication operations per timestep. Running LQG control is therefore computationally inexpensive, achieving optimal performance with computations that can be carried out within a fraction of the system's frame delay.

\subsection{Model building for LQG in an AO context}

\begin{table}
    \centering
    \begin{tabular}{|c|c|c|}
        \hline
        \textbf{Matrix} & \textbf{Purpose} & \textbf{Dimension}\\\hline
        $A$ & State-to-state time evolution & $s \times s$\\
        $B$ & Input-to-state time evolution & $s \times p$\\
        $C$ & State-to-measurement translation & $m \times s$\\
        $D$ & Input-to-measurement translation & $m \times p$\\
        $W$ & Process noise covariance & $s \times s$\\
        $V$ & Measurement noise covariance & $m \times m$\\
        $Q$ & State weighting in cost (not tuned here) & $s \times s$\\
        $R$ & Input weighting in cost (not tuned here) & $p \times p$\\
        $S$ & State-input correlated cost (not tuned here) & $p \times s$\\
        $P_\obs$ & Covariance of Kalman filter state estimate & $s \times s$\\
        $P_\con$ & Inner product kernel for optimal control cost & $s \times s$\\
        $K$ & Kalman gain: optimal correction for state estimates & $s \times m$\\
        $L$ & LQR gain: optimal function of state to use as input & $p \times s$\\\hline
    \end{tabular}
    \caption{Summary of matrices in LQG control, with state size $s$, measurement size $m$, input size $p$.}
    \label{tab:matrices}
\end{table}

Table~\ref{tab:matrices} describes all of the matrices involved. To apply LQG control to adaptive optics, it suffices to specify $A, B, C, D, W, V$ describing the system's time-evolution. 

We follow a model choice similar to that in work by Meimon et al.\cite{Meimon}, in which tip-tilt dynamics are the sum of vibration and atmospheric components, as described using autoregressive models of the form

\begin{align}
    x_i = \sum_{k=1}^p a_k x_{i-k} + \epsilon_i,
\end{align}

where $\epsilon_i$ is i.i.d Gaussian noise and $a_k$ are known coefficients. Here, we restrict models to the AR(2) case, with $x_i = a_1 x_{i-1} + a_2 x_{i-2} + \epsilon_i$. Identification consists of analyzing an open-loop timeseries in order to find a combination of autoregressive models that describe its dynamics well.

To identify the system, we first take the power-spectral-density (PSD) of the open-loop timeseries. The PSD is computed using Welch's method with a Hanning window of size $2^{\left\lfloor \log_2 \frac{N}{4}\right\rfloor}$ for a timeseries of length $N$, and with the DC component ($f = 0$) removed. We perform a linear fit in log-log space in a frequency range $\qty[\frac{f_s}{120}, \frac{f_s}{3}]$. We mask out any datapoints that are greater than 3 standard deviations, replacing them with a new log-log linear interpolation, to get the modified PSD we use to fit the atmospheric model.

We identify up to $N_{\text{vib,max}} = 10$ vibrational modes in each of tip and tilt by iteratively finding the largest peak in the PSD, using the \textit{scipy.signal.find\_peaks}\cite{scipy} function, and again ensuring that it falls within the frequency range $\qty[\frac{f_s}{120}, \frac{f_s}{3}]$. For each one, we make a modified PSD, where all values that are 3 standard deviations above the linear fit are masked out, except those around the identified central frequency. This provides a clear peak that can be well described by an AR(2) model.

To fit an autoregressive model, we use the PSDs rather than the timeseries so that we can exploit spectral separation between components, as well as separating out measurement noise, which tends to dominate above $\frac{f_s}{3}$. We use the bounded Levenberg-Marquardt minimization method as implemented in the Python \textit{lmfit}\cite{lmfit} package to fit the log of a modified PSD to an autoregressive power-law model of order 2.\cite{ox}

\begin{align}
    P(f) = \frac{\sigma^2 / f_s}{\abs{1 + a_1 \exp(-i 2\pi f / f_s) + a_2 \exp(-i 4\pi f/f_s)}^2}\\
    \ln P(f) = 2\ln \sigma - \ln f_s - 2 \log \abs{1 + a_1 \exp(-i 2\pi f / f_s) + a_2 \exp(-i 4\pi f/f_s)}.
\end{align}

For the atmospheric trend, we fit an AR(2) model as above (fitting for $a_1, a_2, \sigma$), which has been noted by Meimon et al.\cite{Meimon} as providing an adequate description of atmospheric turbulence; however, this order could be increased if necessary to capture more complex dynamics that are correlated in time. For each vibration peak, we fit the same model but for $f, k, \sigma$, where $a_1, a_2$ are related to $f, k$ by discretization of a damped exponential:

\begin{align}
    \begin{split}
    a_1 &= 2 e^{-k2\pi (f/f_s)} \cos(2\pi \sqrt{1 - k^2} (f/f_s))\\
    a_2 &= -e^{-4\pi k (f/f_s)}
    \end{split}
\end{align}

$f$ is provided as an initial value from the peak fit, rather than a fixed parameter, since its true value may be different than the peak-fitted value by up to $\Delta f = \frac{f_s}{2^{\left\lfloor \log_2 \frac{N}{4}\right\rfloor + 1}}$ due to the window size and the Nyquist sampling theorem. It is bounded to within $\pm \Delta f$ of its initial value. 

Finally, we identify the measurement noise by averaging the portion of the PSD in the frequency range $\qty[\frac{f_s}{3}, \frac{f_s}{2}]$ (the left endpoint of which may be adjusted: the selected portion of the PSD should be relatively flat, so that there are no dynamics being included in the estimate) and taking the root-mean-square estimate,

\begin{align}
    \sigma_v = \sqrt{\langle f_s P_{f \geq f_s/3}\rangle},
\end{align}

which accurately represents measurement noise as it recovers the average Fourier transform, which sends a Gaussian in time to one in frequency.

Having found the autoregressive coefficients and measurement noise, we build the system matrices. An AR(2) model with process noise $\epsilon_i \sim \mathcal{N}(0, \sigma^2)$ can be described in state space by

\begin{align}
    \begin{split}
        \vec{x}[k] = \mat{x_k & x_{k-1}}^\intercal,\ 
        A = \mat{a_1 & a_2 \\ 1 & 0},\ 
        B = \mat{0 & 0}^\intercal,\
        C = \mat{1 & 0},\
        W = \mat{\sigma^2 & 0 \\ 0 & 0}
    \end{split}
\end{align}

and to combine models, we stack their $A$ and $W$ matrices in block-diagonal form, while concatenating their $B$ matrices vertically and their $C$ matrices horizontally. This is combined with $D = \mat{1}$ and $V = \mat{\sigma_v^2}$ independently of the models being used. In total, a system matrix for both modes that models atmospheric turbulence and one vibrational mode has the representation (where all blank entries are implicitly zeros),
        
\begingroup
\allowdisplaybreaks
\begin{align*}
        \vec{x}_k &= \mat{x^{tip,atm}_{k} & x^{tip,atm}_{k-1} & x^{tip,vib}_{k} & x^{tip,vib}_{k-1} & x^{tilt,atm}_{k} & x^{tilt,atm}_{k-1} & x^{tilt,vib}_{k} & x^{tilt,vib}_{k-1}}\\
        A &= \mat{a^{tip,atm}_1 & a^{tip,atm}_2 & & & & & & \\ 1 & 0 & & & & & & \\ & & a^{tip,vib}_1 & a^{tip,vib}_2 & & & & \\ & & 1 & 0 & & & & \\ & & & & a^{tilt,atm}_1 & a^{tilt,atm}_2 & & \\ & & & & 1 & 0 & & \\ & & & & & & a^{tilt,vib}_1 & a^{tilt,vib}_2 \\ & & & & & & 1 & 0}\\
        B &= \mat{\mathbf{0}}_{8\times 2}\\
        C &= \mat{1 & 0 & 1 & 0 & & & & \\ & & & & 1 & 0 & 1 & 0}\\
        D &= \mat{1 & 0 \\ 0 & 1}\\
        W &= \diag(\mat{\sigma_{tip,atm}^2 & 0 & \sigma_{tip,vib}^2 & 0 & \sigma_{tilt,atm}^2 & 0 & \sigma_{tilt,vib}^2 & 0})\\
        V &= \mat{\sigma_{v,tip}^2 & 0 \\ 0 & \sigma_{v,tilt}^2}.
\end{align*}
\endgroup

\subsection{Delay transform}

Adaptive optics control loops typically have a frame delay, which is characterized in the model by changing the update rule to

\begin{align}
    \vec{x}[k+1] = A\vec{x}[k] + Bu[k-d],
\end{align}

for a delay of $d \geq 0$ frames. In the state-space formulation, this can be done by augmenting the state as follows. Let $\vec{x}' = \mat{\vec{x} & \vec{u}_{k-1} & \dots \vec{u}_{k-d}}^\intercal$ be the new state, with dimension $n + pd$. The corresponding system matrices are updated as follows:

\begingroup
\allowdisplaybreaks
\begin{align*}
    A' &= \mat{A_{n\times n} & \vec{0} & \vec{0} & \cdots & \vec{0} & \vec{0} & B_{n \times p} \\ \vec{0} & \vec{0} & \vec{0} & \cdots & \vec{0} & \vec{0} & \vec{0} \\ \vec{0} & I_{p\times p} & \vec{0} & \cdots \vec{0} & \vec{0} & \vec{0} \\ \vdots & \vdots & \vdots & \ddots & \vdots & \vdots & \vdots \\ \vec{0} & \vec{0} & \vec{0} & \cdots & I_{p\times p} & \vec{0} & \vec{0} \\ \vec{0} & \vec{0} & \vec{0} & \cdots & \vec{0} & I_{p \times p} & \vec{0}}_{(s + pd) \times (s + pd)}\\
    B' &= \mat{\vec{0}_{n \times p} & I_{p \times p} & \vec{0} & \dots & \vec{0} & \vec{0}}^\intercal_{(n + pd) \times p}\\
    C' &= \mat{C & \vec{0} & \vec{0} & \dots & \vec{0} & D}_{m \times (s + pd)}\\
    D' &= \mat{0}_{m \times p}\\
    W' &= \mat{W & \vec{0} \\ \vec{0} & \vec{0}_{pd \times pd}}\\
    V' &= V
\end{align*}
\endgroup

\subsection{Method validation through simulations}

We validate LQG control and the effects of introducing delays first on a toy problem, then on data that are representative of real tip-tilts. For both of these simulations, we generate sets of process noise $\vec{w}[k]$ and measurement noise $\vec{v}[k]$, and iterate $\vec{x}[k+1] = A\vec{x}[k] + B\vec{u}[k] + w[k]$ and $\vec{y}[k] = C\vec{x}[k] + D\vec{u}[k] + \vec{v}[k]$ for various choices of control law $\vec{u}_k(\vec{y}_1, \dots, \vec{y}_k)$. Direct comparisons are possible due to the same $\vec{w}_k, \vec{v}_k$ being used.

For an identity model (i.e. the state remains constant except for process noise) with a frame delay, which is described by the state-space system $A = \mat{1 & 1 \\ 0 & 0}, B = \mat{0 & 1}^\intercal, C = \mat{1 & 0}, D = \mat{0}, W = \mat{1 & 0 \\ 0 & 0}, V = \mat{1}$, we simulate four control laws: open-loop, where $\vec{u}_k = \vec{0}$; integrator control, where $\vec{u}_k = -g\vec{y}_k$ given some constant gain $g$, in this case 0.1; LQG control for the equivalent system without a frame delay; and LQG control for this system. 

\begin{figure}
    \centering
    \includegraphics[width=\textwidth]{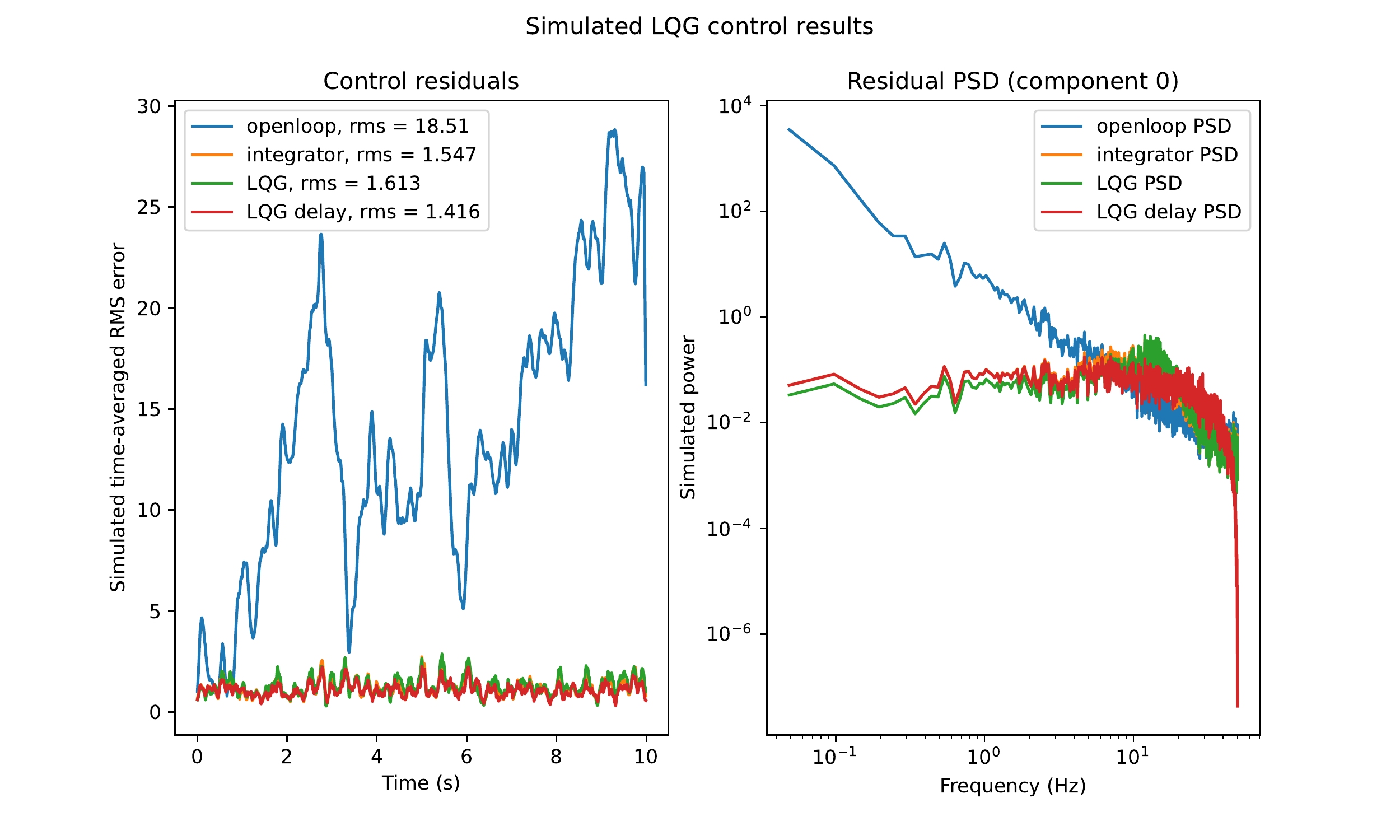}
    \caption{Simulated LQG control with and without delay. On the open-loop data (blue), control is simulated using three schemes: the integrator (orange), LQG without the delay terms (green), and LQG with the delay terms (red). We see LQG with the delay terms performs better than the others.}
    \label{fig:simlqg_delay}
\end{figure}

Figure~\ref{fig:simlqg_delay} shows the results. As expected, we see that LQG control for the true system provides the best performance, and further that the controlled residuals are spectrally flat.

\begin{figure}
    \centering
    \includegraphics[width=0.8\textwidth]{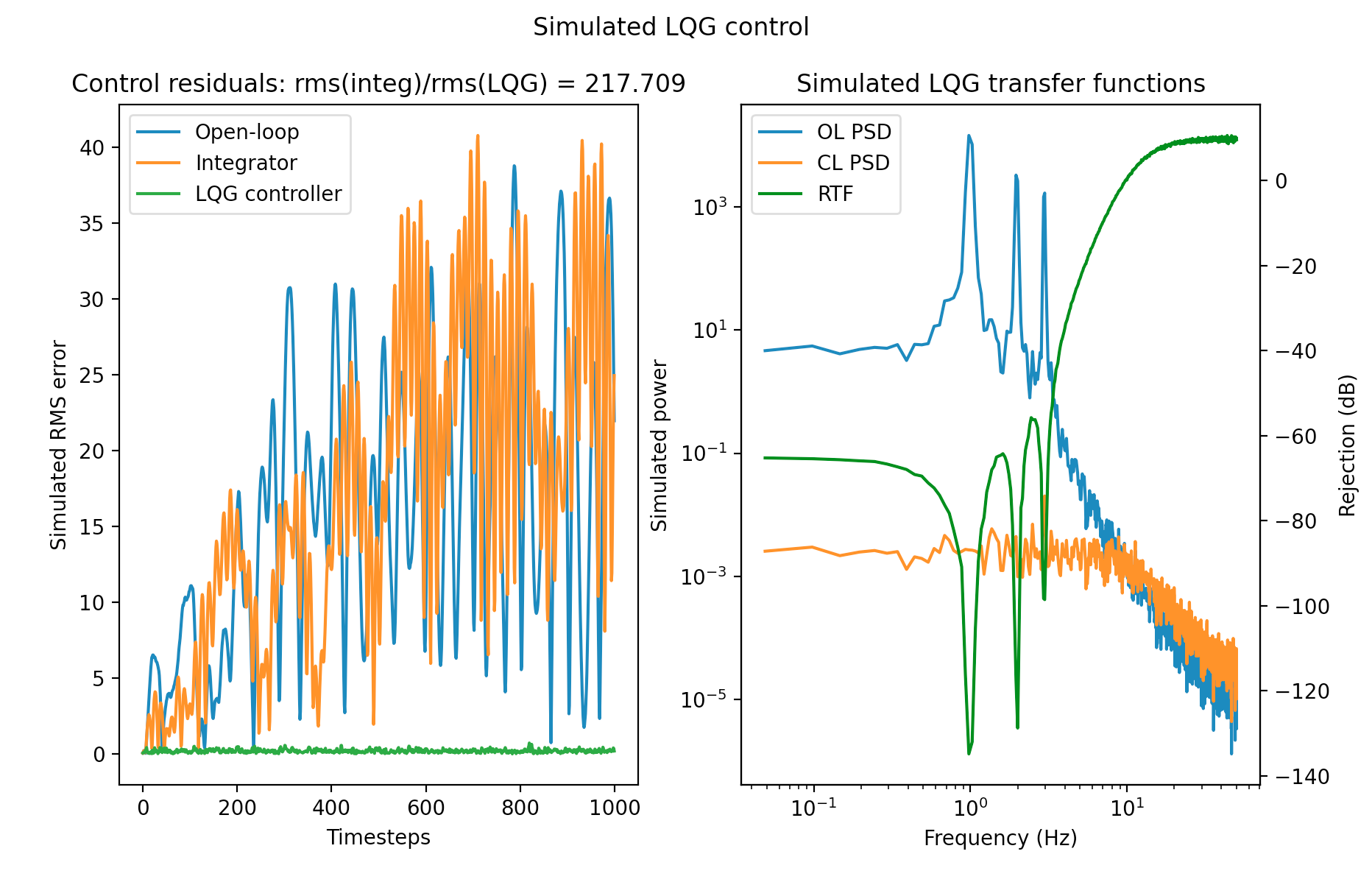}
    \caption{Simulated LQG control for three-vibration signal}
    \label{fig:simlqg_threevib}
\end{figure}

We further test LQG against integrator control with a model containing three vibration peaks, the results of which are shown in Figure~\ref{fig:simlqg_threevib}. As expected, the LQG residuals are flat both in time and frequency, and the empirical rejection transfer function (closed-loop PSD divided by open-loop PSD) shows strong inverse peaks at each of the vibration peaks.

\section{Setup and Experiments}

\subsection{Lab setup}

We made use of the FAST\cite{fast_orig} (Fast Atmospheric Self-coherent camera Technique) method on the SEAL\cite{SEALBecky,FASTBen} (Santa cruz Extreme Adaptive optics Laboratory) testbed for this work. FAST was implemented using an Andor Zyla 5.5 sCMOS camera and an ALPAO 97 actuator deformable mirror, with the intermediate wavefront sensing steps carried out within the Python codebase described in \S\ref{sec: software_setup}.

Instead of generating turbulence on the bench, we simulate disturbances by applying tip and tilt commands in closed loop that are then corrected. This has the advantage of allowing us to define disturbances that precisely match the dynamic model of the LQG controller, or that deviate from it in known ways.

\subsection{Calibration and measurement}

Before experiments can be run, several calibration steps are executed. These steps ensure the optics are aligned and will produce the desired results in simpler test cases. 

The first step is optical flattening, which is described in detail in a separate work by Gerard et al.\cite{FASTBen}, but also summarized here. This works by optically flattening the ALPAO DM with a Thorlabs Shack Hartmann Wavefront Sensor (model WFS-20). We use a classical poke matrix, singular value decomposition (SVD)-based psuedo inverse to generate the command matrix while minimizing the propagation of unwanted DM modes such as waffle, and then iteratively closed the loop to optically flatten the ALPAO DM, typically converging at $\sim$40 nm rms. We repeat this process $\sim$daily in order to mitigate known ALPAO quasi-static drift effects. 

After flattening, optical alignment is run. This process consists of a tip/tilt grid search to find the optimal best-flat position of the camera image. In order to find this optimal position, we look at the modulation transfer function (MTF) of the image at each candidate position, possibly after manually steering in order to reach a region where interference effects are clearly visible. The MTF is compared against a pre-computed ideal MTF, which consists of a central lobe surrounded by two side lobes, by computing the contrast ratio of a side lobe to the central lobe. We run this grid search once over a coarse grid and once over a fine grid within the optimal region. Indicators of reaching an optimal position are clearly visible and separated side lobes in the MTF, and a contrast ratio heatmap that peaks roughly in the center of the grid. This optimal position is saved as the best flat, and an averaged image taken at this position is saved for use in measurements.

From an aligned position, an interaction matrix is computed, by applying ideal Zernike shapes and noting the change relative to the best flat in $I_-$ space. This results in a reference vector, whose outer product with itself is the interaction matrix. The SVD-based pseudo-inverse of this matrix is the command matrix, which we use to measure Zernike coefficients from images. The SVD cutoff to compute the command matrix is optimized to account for changing day-to-day conditions in the lab. The Zernike modes in an image are then measured by computing the change in $I_-$ space relative to a flat reference, as above, and projecting the result onto the basis of ideal Zernike responses in the command matrix. 

\begin{figure}
    \centering
    \includegraphics[width=\textwidth]{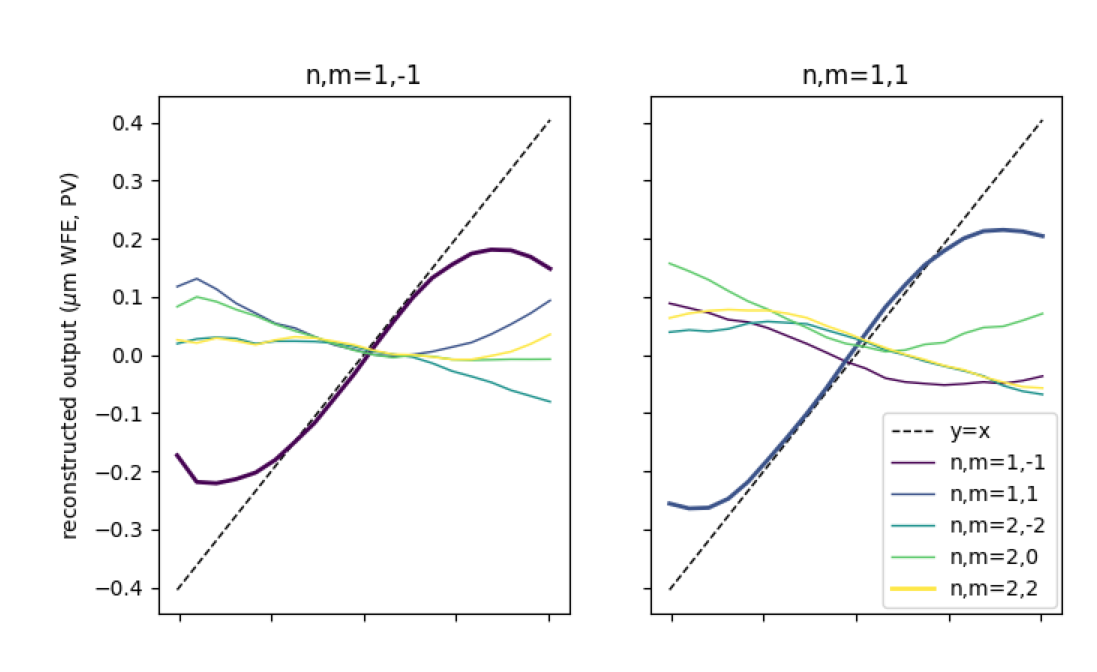}
    \caption{Linearity of FAST in each Zernike mode}
    \label{fig:linearity}
\end{figure}

We check system linearity by applying linear perturbations in each Zernike mode, and plotting against the actual measured output in each mode. Under ideal conditions, this graph is close to $y = x$ in the mode being perturbed, i.e. a unit input results in a unit output, and close to $y = 0$ in other modes, i.e. no cross-talk between modes. This holds until the perturbation is sufficiently off the focal plane mask, significantly decreasing the intensity of interference fringes. If these conditions are not visibly satisfied, the above steps are rerun. Figure~\ref{fig:linearity} shows representative linearity plots on the bench.

\subsection{Software setup}
\label{sec: software_setup}
A SEAL adaptive optics experiment is completely specified using the system optics (the actual hardware or a simulation mode), the duration of an experiment, and the disturbance schedule (specifying the tip and tilt disturbance values to be put on at each timestep). This is paired at runtime with a controller, which is a function that takes in a system measurement and returns the tip and tilt output to apply. In this way, multiple controllers can be run on the same random process in simulation, and can be run sequentially on the bench.

Before running an experiment, we check hardware alignment by ensuring that no component of the Zernike measurement at the best-flat DM position exceeds a threshold (arbitrarily set at $10^{-3}$ of the maximum possible actuator input). If this is the case, alignment is automatically rerun and the command matrix is renewed. If this automatic alignment fails after a few iterations, the experiment terminates, and we attempt manual realignment. 

\begin{figure}
    \centering
    \includegraphics[width=0.8\textwidth]{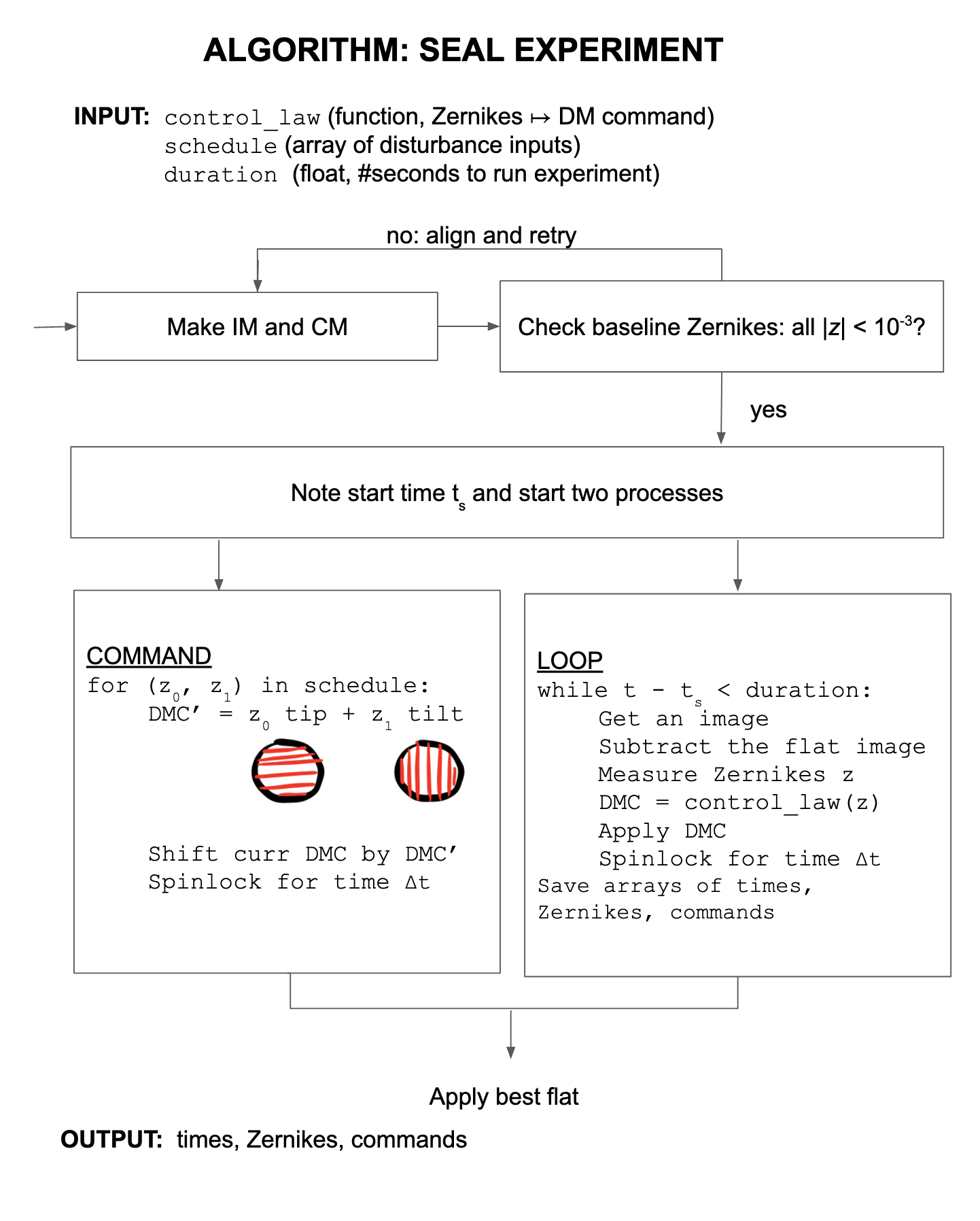}
    \caption{Computational flow of a SEAL adaptive optics control experiment.}
    \label{fig:seal_experiment}
\end{figure}

After checking alignment, the control loop is run by starting two independent processes, as shown in Figure~\ref{fig:seal_experiment}. The first generates a schedule of open-loop disturbances in tip and tilt to apply at each timestep, and applies them to the DM at the appropriate times. The second runs the control loop: in each iteration, it gets an image from the camera, processes it to generate the measured Zernike coefficients, computes the command based on that measurement, and applies it to the DM. The second process also logs the times at which each event happens, the measured coefficient values, and the control commands applied.

\subsection{Control loop delays}

Although Python is not a language typically used for real-time control,  we developed a solution, described below, to enable sufficient high-speed performance. This solution combined with the preferred user interface and modular abilities of Python made this approach the best choice for the purposes of our laboratory demonstration.

A key aspect to running real-time control in Python is achieving concurrent execution, so that the delay between a disturbance and its correction is either small enough to be negligible, or a relatively constant multiple of the frame rate so that the controller can compensate for it. This is achieved in our implementation by spin-locking each thread, forcing it to stay in a dummy infinite loop until the system time reaches the time of the previous frame plus $\Delta t$. This works in conjunction with Python \textit{multiprocessing.Process} objects, which run separate instances of Python and therefore avoids interdependences that would otherwise interfere with the execution order due to the Global Interpreter Lock.\cite{gil}

To account for slight imprecision in the timing controls in Python, a SEAL experiment will drop a frame if the loop computation time ever exceeds the loop rate. For instance, if a loop iteration was set to end at $t = 0.01$s, but the iteration takes $0.013$s to complete, the experiment will spinlock until $t = 0.02$s so that the system matches the underlying discrete-time model. In practice, at most one frame is lost per 1000-frame experiment and most control experiments do not drop any frames in this way, so overall control performance is not significantly affected by the implementation.

\begin{figure}
    \centering
    \includegraphics{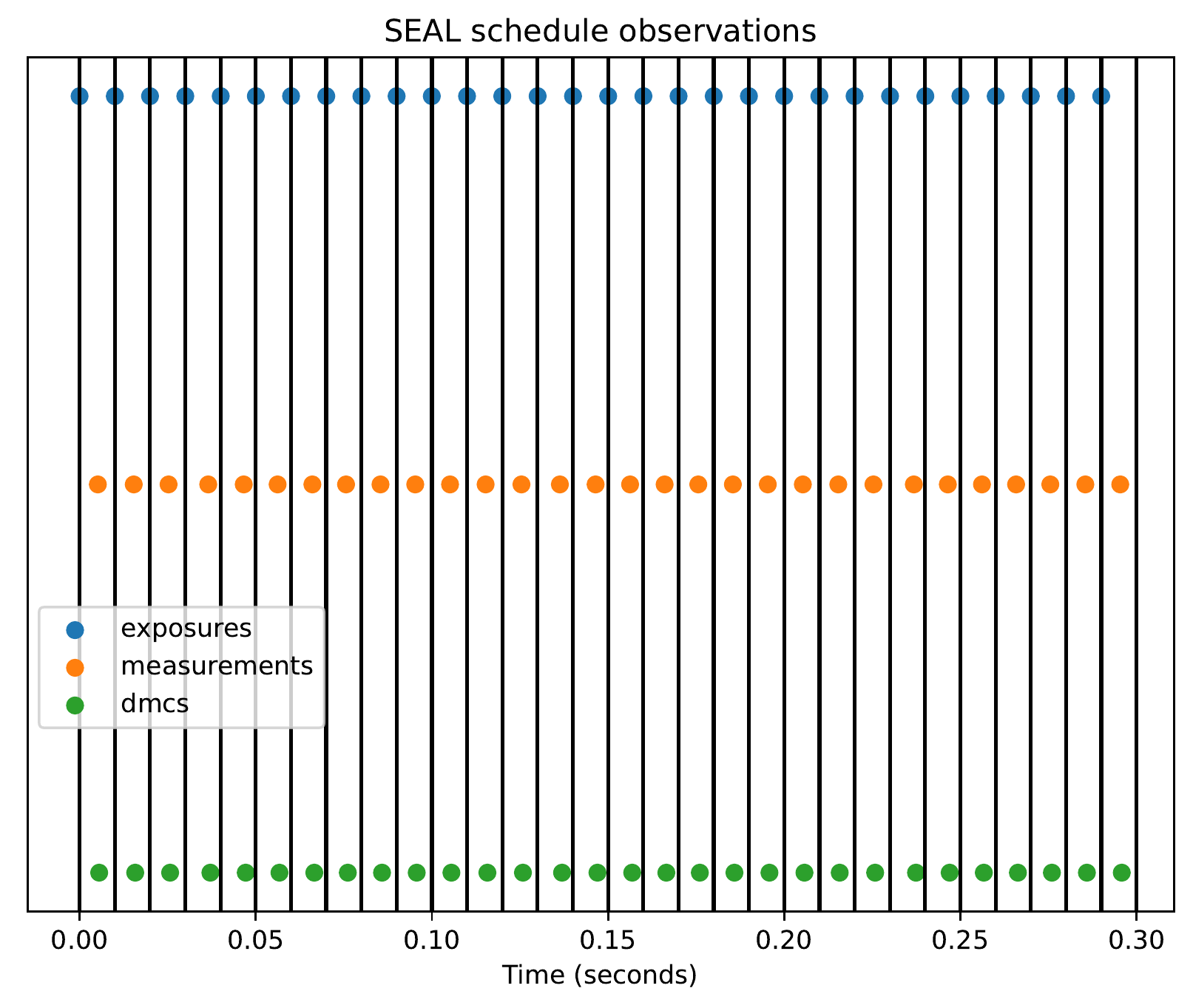}
    \caption{SEAL schedule in closed loop. Each line is the extent of a planned frame, which starts with an exposure (in blue), then a measurement (in orange), then a DM command (in green). For the controller to be in sync, all three dots should appear before the next vertical line.}
    \label{fig:seal_schedule}
\end{figure}

Figure~\ref{fig:seal_schedule} shows a schedule, as measured on the bench, for a typical run of this method. We note that each corrective DM action is applied within half a frame of the corresponding camera exposure, and the loop runs at 100 Hz without any events occurring out of sync or drifting over time. This shows us that the experimental setup is robust enough to satisfy the assumptions of LQG control.

\begin{figure}
    \centering
    \includegraphics[width=\textwidth]{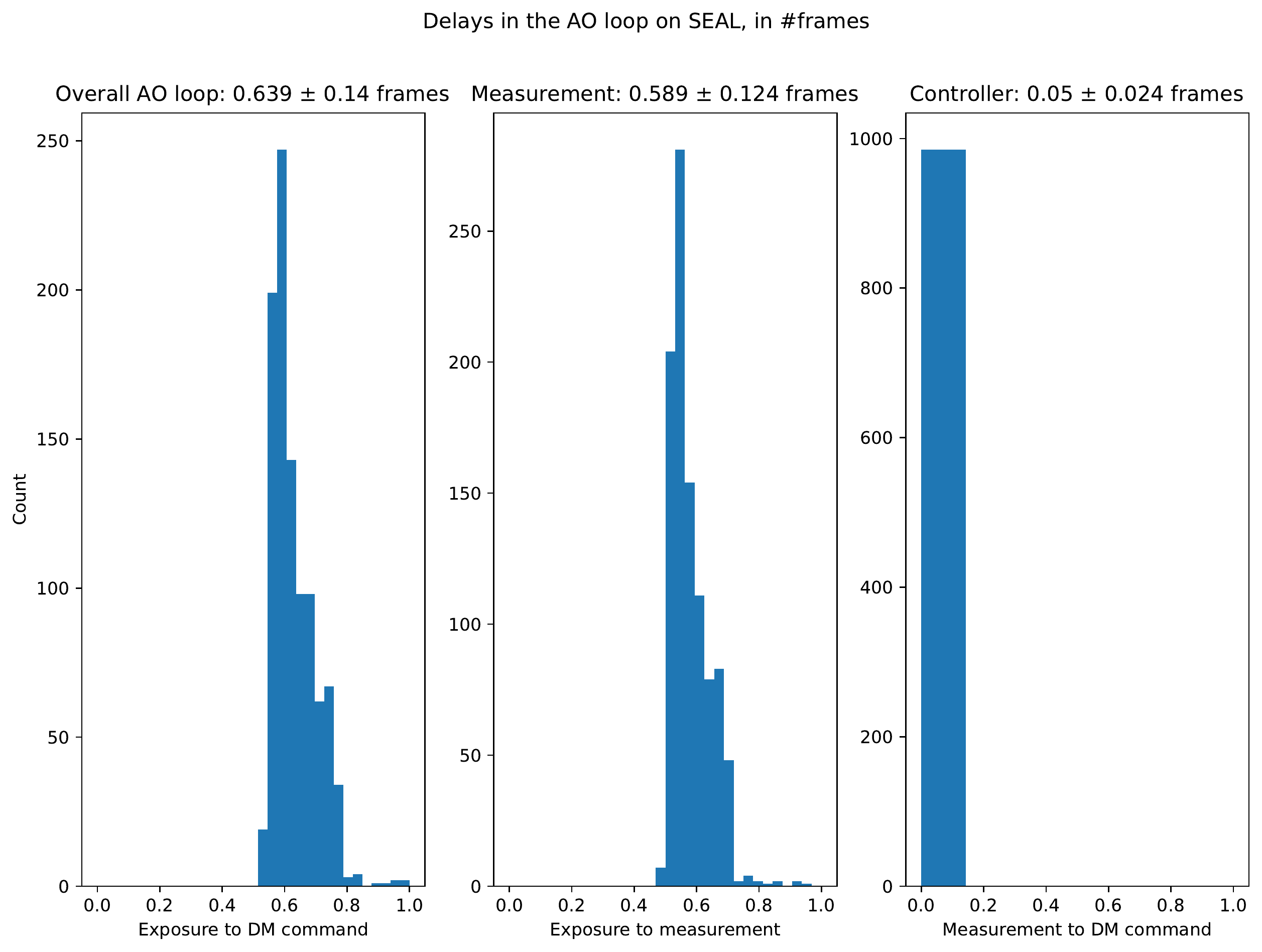}
    \caption{SEAL delay histograms. From left to right, the figure depicts the time from exposure to DM command, exposure to measurement, and measurement to DM command.}
    \label{fig:seal_delays}
\end{figure}

We can further see the effectiveness of this implementation by looking at the delays between consecutive events in the loop. Histograms of these delays for the same typical bench run (lasting 10 seconds/1000 frames) and normalized to 1 frame = 0.01 seconds are shown in Figure~\ref{fig:seal_delays}. We note that the control computation is relatively constant, as we would expect from LQG control (since it consists of the same few matrix multiplications in each iteration), whereas the measurement time is slightly longer but still relatively constant. Since all operations run well within 1 frame, barring a negligible number of dropped frames as described above, we conclude that the bench implementation of closed-loop control is robust.

\section{Results, Discussion, and Future Work}



In order to run an LQG control experiment, we first take a 100-second open-loop timeseries. On-sky, this would be equivalently achieved by periodically performing a pseudo-open-loop reconstruction. We use this open-loop time series to build an LQG controller as previously described. For the purposes of control experiments, we specify the AR order of the atmospheric model and the number of vibration peaks to be used in the identification, to ensure that the identified model is able to perfectly match the underlying model used to make the disturbance; on-sky, this tuning would be unnecessary, and we would set these to 2 and 10 respectively.

\begin{figure}
    \centering
    \includegraphics[width=0.8\textwidth]{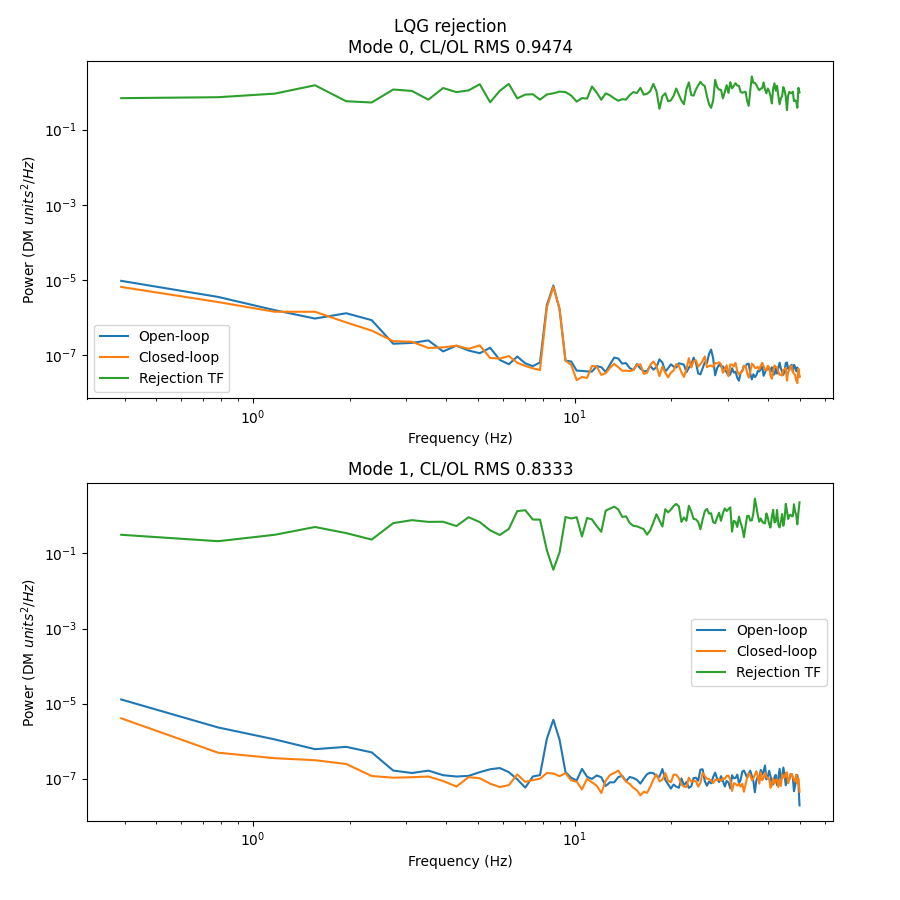}
    \caption{SEAL LQG results in the frequency domain}
    \label{fig:seal_lqg}
\end{figure}

Figure~\ref{fig:seal_lqg} shows the open- and closed-loop PSDs, and the rejection transfer function (their quotient), in both tip and tilt. As expected, it notches out the induced vibration peak, but only in one mode; this may be due to imperfect linearity observed in the other mode, which we discuss further in \S\ref{sec: future_work} along with additional possible future work to improve laboratory performance.

\label{sec: future_work}
From Fig. \ref{fig:linearity}, the FAST linear range is about 0.1 $\mu$m PV, converting to 3.4$\times10^{-6}$ (ALPAO DM units)$^2$. Looking at Fig. \ref{fig:seal_lqg}, this linear range is below amplitudes measured at frequencies $\lesssim$1 Hz, which means FAST non-linearities should begin to dominate at these low temporal frequencies, potentially explaining the limited closed-loop gains as we have currently measured. Similarly, the Mode 0 vibration peak centered around 8 Hz, reaches amplitudes above this non-linear regime, whereas the Mode 1 vibration peak shows smaller amplitudes, potentially explaining why the latter mode is corrected but the former is not. A future demonstration where open loop vibration is introduced with the DM at controlled amplitudes within the FAST linear range could help improve performance.

Two additional major ways in which these experiments could be extended would be by increasing the loop rate by identifying and removing computational bottlenecks, and by testing more computationally intensive control algorithms. 

The main computational bottleneck in closed-loop control using FAST is the measurement step, which involves a 2D Fourier transform, multiplication in Fourier space by a mask, and an inverse Fourier transform back. This computation takes about 3-5ms on SEAL, and would ideally take $<$1ms. The mask is sufficiently large that an approach using the fast Fourier transform is faster than computing individual components, and the camera dimension is sufficiently large that Fourier methods are faster than spatial-domain convolution. However, since the measurement computation is a static operation over time (consisting of Fourier transforms of the same dimension and a multiplication with the same mask at each iteration), using a lower-level language or offloading some of the computations to a GPU would provide a substantial speedup; in particular, this would make loading the entire image array into Python unnecessary, instead only requiring the much-smaller array of Zernike coefficients to be loaded. Further, the overall operation is linear, suggesting that it could be even further condensed into a single matrix multiplication. A lower-level language would also be well suited to more tightly controlling event timings. In particular, Rust is a good candidate for this task, as its binaries can be loaded as Python libraries and it allows for low-level control and optimization while ensuring memory safety. Porting some or all of the underlying optics and experiment code to Rust would therefore still allow the control algorithms, or other aspects that may require customization, to be written in Python.

This work serves as a proof of concept for testing arbitrarily computationally intensive control algorithms in this setup provided they are delay-tolerant. Due to concurrent processing, the only limiting factor on a control law is the loop rate, and control laws that take longer than the loop rate to compute can still be run, provided they are designed to tolerate this inherent delay. This allows for testing of algorithms like predictive wavefront control\cite{predictive} with FAST, using prototype models in Python. This lowers the barrier between simulation and bench testing of a control algorithm.

Delay tolerance can be extended further to the controller model. The vibration model used in the LQG controller is a discrete-time realization of a continuous-time process, so by using the underlying infinitesimal generator matrix to time-evolve the underlying model for arbitrary lengths of time, an externally-imposed loop rate could be relaxed. The atmospheric model is based on discrete-time computations, but could be approximated in continuous time via linear interpolation. In the case where the lag is Markovian, which is a reasonable assumption in the vibration LQG case where the problem complexity can increase or decrease noticeably with the changing dimensions due to different numbers of peaks being identified, optimal observers and controllers similar to those used in this work\cite{StochasticJump,KalmanContinuous} can be designed. This would allow the controller to function better in situations where the loop delay is non-integer or difficult to characterize.

\section{Conclusion}


We have demonstrated the effectiveness of LQG control for tip-tilt using FAST, and developed robust testing infrastructure for real-time wavefront control. Our main findings are as follows.
\begin{itemize}
    \item We present a laboratory demonstration of LQG control with FAST, showing RMS wavefront error improvement compared to an integrator (over all temporal frequencies) and frequency-domain notching of the injected and corrected 9 Hz vibration mode. This is an important milestone for demonstrating the promise of using such LQG technology to suppress vibrations at moderate- to high-speeds with a focal plane WFS.
    \item In these tests we demonstrate the benefit of including plant dynamics into the LQG model, primarily through the inclusion of a delay term.
\end{itemize}
These results serve as proof of concept for the use of high-speed focal plane WFSs in control experiments, particularly where the control algorithm has high computational requirements. We have also made use of a pure Python modular codebase, ensuring that varying the control law or features of the control loop for future research is straightforward. The modular structure further ensures that the control loop could be further optimized, possibly increasing the loop rate, without changing the Python interface for defining the control law. 
\acknowledgements
AS thanks Deana Tanguay and Enrico Ramirez-Ruiz for support throughout the Lamat REU program, and Parth Nobel and Aled Cuda for consultations on implementing concurrent execution. 

\appendix

\section{Overview of FAST}
\label{sec: fast_rev}
\begin{figure}[!h]
\centering
\includegraphics[width=0.98\textwidth]{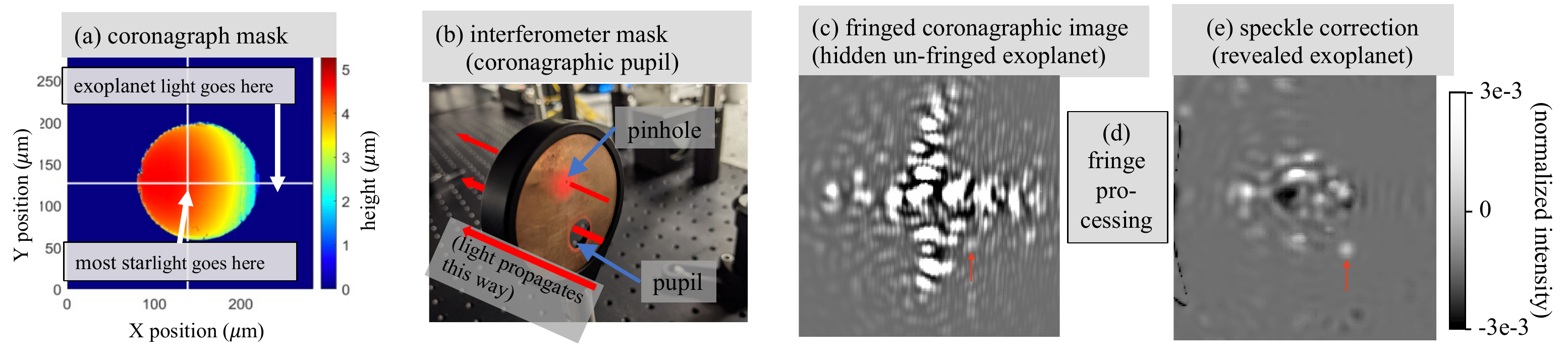}
\caption{Overview of FAST, adapted from Ref. \citenum{FASTBen}, using images from the Santa Cruz Extreme AO Laboratory (SEAL) high contrast testbed\cite{SEALBecky}. A custom coronagraph mask (a) is designed to both (i) attenuate starlight while transmitting exoplanet light, and (ii) enhance the stellar fringe visibility using an interferometer mask in the downstream coronagraphic pupil (b). This design enables the detection of an interference pattern on top of the leftover stellar speckles in the coronagraphic image (including for low order modes that are blocked by the coronagraph mask in panel a but still transmitted through the pinhole in panel b), even when only a few photons are detected per pixel (c), enabling coronagraphic science imaging and AO wavefront sensing. The fringed stellar speckles are then processed (d) and further attenuated using both a deformable mirror and via post-processing  (e), revealing an un-fringed exoplanet.}
\label{fig: fast_sum}
\end{figure}
Fig. \ref{fig: fast_sum} shows an overview of the Fast Atmospheric Self-coherent camera (SCC) Technique (FAST)\cite{fast_orig}, adapted from Ref. \citenum{FASTBen}. FAST relies on the SCC technique\cite{scc_orig}, in which coherent starlight is interfered with itself to enable a measurement and correction of the complex speckle electric field (i.e., phase and amplitude) in the coronagraphic image. Reviewing a mathematical description of the SCC/FAST image, the recorded image on the focal plane detector, $I$, produced from noiseless (i.e., without photon/detector/sky background noise) propagation of wavefront error (WFE; i.e., from atmospheric and/or instrumental origin) through to the detector, can be described by
\begin{equation}
    I=I_S\{\theta\}+I_P\{\theta\}+I_R\{\theta\}+2\sqrt{I_S\{\theta\} I_R\{\theta\}}\;\mathrm{cos}[2\pi(\xi_0/\lambda)\phi\{\theta\}],
    \label{eq: scc}
\end{equation}
where $\theta$ represents the two-dimensional coordinates in the image, $I_S$ is the un-fringed stellar speckle intensity component, $I_P$ is the un-fringed planet intensity component, $I_R$ is the pinhole point-spread-function (PSF) component (i.e., the PSF recorded if the main pupil aperture was blocked), and $\phi$ is the electric field phase difference between the $I_S$ and $I_R$ image components. The spatial scale of the fringe term $2\sqrt{I_S I_R}\mathrm{cos}[2\pi(\xi_0/\lambda)\phi]$ is set by the Lyot stop's pupil-to-pinhole separation ($\xi_0$, projected onto the entrance pupil) and observing wavelength, $\lambda$, enabling sub-$\lambda/D$ spatial scales (between $\lambda/[0.5 \xi_0]$ and $\lambda/[1.5 \xi_0]$) to be be isolated in the Fourier plane of the image without bias from the un-fringed exoplanet signal $I_P$. The phase of a given stellar speckle, which is not normally measureable in a single coronagraphic image (i.e., not encoded in $I_S$ or $I_R$), is encoded by the relative position of fringes on that speckle, which changes with $\phi$. This relative phase can be isolated in the Fourier plane of the image (known as the optical transfer function or OTF, the amplitude of which is known as the modulation transfer function or MTF) via a binary mask multiplication of these ``side lobe'' spatial frequencies; the complex-valued image plane after Fourier-filtering the MTF sidelobe is commonly referred to as $I_-$, representing the fringe amplitude and phase components in the same plane as the detector. This phase measurement is calibrated empirically by applying sine and cosine waves over the full spatial frequency range of the DM (or for the purposes of this paper, analogously with low order Zernike modes). By recording images (i.e., eq. \ref{eq: scc}) that are then converted to $I_-$ during this calibration routine to build an AO-based command matrix, DM commands can then be generated from a single image with an unknown speckle pattern pattern via a linear least-squares-based matrix vector multiplication (MVM; ideal for closed-loop AO operations if fringes can be detected on millisecond timescales).

FAST's key innovation compared with previous focal plane wavefront sensing and control is the coronagraphic focal plane mask (FPM; Fig. \ref{fig: fast_sum}a), which provides a high enough fringe visibility (i.e., fringed vs. un-fringed components) in the final image such that fringes can be detected on millisecond timescales, even when only a few photons are detected per pixel. The classical SCC also generates fringes in the coronagraphic image to enable focal plane wavefront control\cite{scc_orig}, but with fringe visibilities about 10$^6$ times lower than Fig. \ref{fig: fast_sum}c \cite{fast_orig}, removing any possibility of AO-based wavefront control on millisecond timescales. Also due to the nature of the FAST FPM, which sends the core of the PSF (i.e., low order modes of the entrance pupil wavefront) into the off-axis Lyot stop pinhole, in turn transmitted through to the SCC image, FAST is sensitive to low order modes of the entrance pupil wavefront\cite{fast_spie20} while typically the classical SCC is comparatively not sensitive to such modes\cite{scc_orig}.
\section{Linear-quadratic-Gaussian derivation}
\label{sec:lqg_rev}

Solving the optimal control problem in the infinite-horizon case consists of choosing control inputs $\vec{u}_k$ such that the following cost function is minimized:

\begin{align}
    J(\vec{u}_0, \vec{u_1}, \dots) = \lim\limits_{N \to \infty} \frac{1}{N} \E\qty[\sum_{k = 0}^{n-1} \vec{x}_k^\intercal Q \vec{x}_k + \vec{u}_k^\intercal R \vec{u}_k + 2\vec{u}_k^\intercal N \vec{x}_k],
\end{align}

where $Q$ describes the penalty for state deviations, $R$ describes the penalty for excessive control action, and $S$ denotes a cross term between these two (and $\E$ is the expectation operator). We specialize to the case of minimizing the true measurements, i.e. minimizing $\vec{y}_k^\intercal \vec{y}_k$. We use this formulation, as opposed to the case where $D = \mathbf{0}$ shown in most sources\cite{boyd}, because the adaptive optics control problem consists of controlling a measurable property (the tip and tilt positions) of an underlying process, rather than controlling that process itself. The appropriate model for this case has $B = \mathbf{0}$, which under the usual formulation would not be controllable. This results in our choices for $Q, R, S$:

\begin{align}
    \vec{y}_k^\intercal \vec{y}_k = (C\vec{x}_k + D\vec{u}_k)^\intercal (C\vec{x}_k + D\vec{u}_k) = \vec{x}_k^\intercal \underbrace{C^\intercal C}_{Q} \vec{x}_k + 2\vec{u}_k^\intercal \underbrace{D^\intercal C}_{S} \vec{x}_k + \vec{u}_k^\intercal \underbrace{D^\intercal D}_R \vec{u}_k.
\end{align}

This has the advantage of providing a clear choice for these weighting matrices rather than leaving them as ad hoc parameters; any deviations from optimal performance are necessarily due to model mismatches or similar errors, rather than choosing the weighting matrices suboptimally. In practice, this helps with accurate system identification.

We can perform linear-quadratic-Gaussian control by making use of the separation principle in control theory. Without losing optimality, the problem can be split into two parts:

\begin{enumerate}
    \item Design of an optimal observer: computing the minimum mean-square-error state estimates $\hat{\vec{x}}_{k+1} \triangleq \E[\vec{x}_{k+1} \mid \hat{\vec{x}}_k, \vec{y}_k]$.
    \item Design of an optimal noiseless regulator: assuming that our observed states $\hat{\vec{x}}_k$ are the true states $\vec{x}_k$, computing $\vec{u}_k = \arg\min\limits_{\tilde{\vec{u}}_k} J(\tilde{\vec{u}}_0, \tilde{\vec{u}}_1, \dots)$.
\end{enumerate}

Both problems can be reduced to the discrete algebraic Riccati equation (DARE), which is a nonlinear matrix equation of the form

\begin{align}
    \label{eq:dare}
    P = A^\intercal PA - (A^\intercal PB + S)(R + B^\intercal PB)^{-1}(B^\intercal PA + S^\intercal) + Q.
\end{align}

We denote ``$P$ solves Equation \ref{eq:dare}'' by $P = \dare(A, B, Q, R, S)$.

Along the lines of similar LQG derivations\cite{boyd}, we show how to reduce the observer and controller problems to the DARE, and how to compute the optimal gain, just for the control problem, since the case we use here (with $D \neq \mathbf{0}$ and with the cost cross term) is relatively uncommon.

Let the cost starting from a state $\vec{x}_0 = \vec{z}$ be 

\begin{align}
    V(\vec{z}) = \min_{\vec{u}_0, \vec{u}_1, \dots} \sum_{\tau = 0}^\infty \vec{y}_\tau^\intercal \vec{y}_\tau = \min_{u_0, u_1, \dots} \sum_{\tau = 0}^\infty \vec{x}_k^\intercal C^\intercal C \vec{x}_k + 2\vec{u}_k^\intercal D^\intercal C \vec{x}_k + \vec{u}_k^\intercal D^\intercal D \vec{u}_k.
\end{align}

We minimize over one control step by splitting the cost into a one-step term, and the cost from there due to the new state. We pick $u_0$ equal to the $w$ minimizing this form of the cost:

\begin{align}
    \label{eq:onestep}
    V(z) = \min_w \qty(\vec{z}^\intercal C^\intercal C\vec{z} + 2\vec{w}^\intercal D^\intercal C\vec{z} + \vec{w}^\intercal D^\intercal D\vec{w} + V(A\vec{z} + B\vec{w}))
\end{align}

We look for a solution such that $V$ is quadratic in $\vec{z}$, i.e. it has the form $V(\vec{z}) = \vec{z}^\intercal P\vec{z}$. Substituting this in to Equation~\ref{eq:onestep}, we get

\begin{align}
    \label{eq:predare}
    \vec{z}^\intercal P\vec{z} = \vec{z}^\intercal C^\intercal C\vec{z} + \min_\vec{w} (2\vec{w}^\intercal D^\intercal C\vec{z} + \vec{w}^\intercal D^\intercal D\vec{w} + \vec{z}^\intercal A^\intercal PA\vec{z} + 2\vec{w}^\intercal B^\intercal PA\vec{z} + \vec{w}^\intercal B^\intercal PB\vec{w}).
\end{align}

The $\vec{w}$ minimizing the right-hand size, which we will take as $\vec{u}_0$, can be found by quadratic minimization:

\begin{align}
    \vec{u}_0 = \arg\min_\vec{w} \vec{w}^\intercal \underbrace{(D^\intercal C + B^\intercal PA)}_{L_1}\vec{z} + \frac{1}{2} \vec{w}^\intercal \underbrace{(D^\intercal D + B^\intercal PB)}_{L_2} \vec{w},
\end{align}

which has the standard solution

\begin{align}
    \label{eq:lqrgain}
    \vec{u}_0 = -(L_2)^{-1}L_1\vec{z} = -(D^\intercal D + B^\intercal PB)^{-1} (D^\intercal C + B^\intercal PA)\vec{z}.
\end{align}

Substituting this back into~\ref{eq:predare} and dropping the leading $\vec{z}^\intercal$ and trailing $\vec{z}$ on each term (the equation has to be true for all $z$, so the corresponding matrix equation is equivalent), we get

\begin{align}
    P = C^\intercal C + A^\intercal PA - (C^\intercal D + A^\intercal PB)(D^\intercal D + B^\intercal PB)^{-1}(D^\intercal C + B^\intercal PA),
\end{align}

and given $P$ satisfying this, we compute the optimal control command according to Equation~\ref{eq:lqrgain}.


\bibliography{report}
\bibliographystyle{spiebib}

\end{document}